# Fermiology via the electron momentum distribution

G. Kontrym-Sznajd

Institute of Low Temperature and Structure Research, Polish Academy of Sciences,
P.O. Box 1410, 50-950 Wrocław 2, Poland

Investigations of the Fermi surface via the electron momentum distribution reconstructed from either angular correlation of annihilation radiation (or Compton scattering) experimental spectra are presented. The basis of these experiments and mathematical methods applied in reconstructing three-dimensional densities from line (or plane) projections measured in these experiments are described. The review of papers where such techniques have been applied to study the Fermi surface of metallic materials with showing their main results is also done.

## 1. Introduction

Electron momentum density $\rho(p)$ in the extended *p*-space is a crucial point in understanding of electronic properties of quantum systems. This density, defined in the following

$$\rho(\mathbf{p}) = \sum_j n_{\mathbf{k}j} \, | \int_{-\infty}^{\infty} e^{-i\mathbf{p}\cdot\mathbf{r}} \psi_{\mathbf{k}j}(\mathbf{r}) d\mathbf{r} |^2 \qquad (1)$$

contains information not only on the occupied momentum states (and hence the Fermi surface, FS) but also on the Umklapp components of the electron wavefunctions $\psi_{kj}(r)$ in the state *k* of *j-th* band. FS characterises the ground state of metallic systems, their transport and magnetic properties and many other phenomena. Conversion from the extended into reduced zone (to get occupation numbers and resulting FS) is described in Chapter 4.3.

Electron density $\rho(p)$ can be determined by measuring either Compton profiles (CP) [1, 2] or angular correlation of annihilation radiation (ACAR) spectra [3, 4], related to $\rho(p)$ either by its double or single integral, so called line and plane projections, respectively. The main difference between these two experiments, described in Chapter 2, consists in the fact that in the Compton scattering one measures electron momentum densities while in the case of ACAR spectra, the electron-positron (*e-p*) momentum densities.

The three-dimensional (3D) function $\rho(p)$ may be "reconstructed" by measuring profiles along various crystallographic directions. The mathematical problem "reconstruction from projections" has a long history, coming into being independently in various scientific fields from radio-astronomy, geology, physics and biology to medical diagnostics. First papers were published by Cormack [5] and by Mijnarends [6] – they found solutions for line and plane projections, respectively. Cormack's theory with his proposal of applications in multiple X-ray tomography and the first X-ray tomograph made in 1972 by G. Hounsfield, revolutionized medical diagnostics – for which both scientists got the Nobel prize in 1979. In that time there was very fast development of both numerous mathematical methods of computerised tomography [7-9] and various techniques of medical diagnostics such as nuclear magnetic resonance (suggested for imaging in 1973) and positron emission tomography (PET, first developed in 1975).

Meanwhile, such a mathematical question was solved generally in 1917 by Radon [10] who considered a real function $\rho(p)$ in the $N$-dimensional space $R^N$ and its integrals over ($N$-1)-dimensional hyperplanes.

$$\hat{R} \cdot \rho(\mathbf{p}) = g(r,\zeta) = \int_{-\infty}^{\infty} \rho(\mathbf{p}) \delta(r - \zeta \cdot \mathbf{p}) d\mathbf{p}. \qquad (2)$$

$\zeta$ is a unit vector in $R^N$ along $r$ and $r$ is the distance of the ($N$-1)-dimensional hyperplane from the origin of the coordinate system. $N=3$ and $N=2$ corresponds to the reconstruction of 3D densities from plane projections and 2D densities from line projections, respectively. So, to use a solution of the Radon transform for $N=2$, line projections must be collected in such a way that the reconstruction of a 3D density is reduced to a set of reconstructions of 2D densities, performed independently on succeeding parallel planes.

All reconstruction techniques can be classified into two categories: $1^0$ - series expansion methods (algebraic techniques, iterative algorithms or optimisation theory methods) [11] and $2^0$ - transform methods [12]. Transform methods consist in the analytical inversion of the Radon transform. These methods, applied to the image reconstruction of momentum densities from 1D and 2D projections, shortly described in Chapter 3, are presented in the paper [13] with references of their applications to study electron (or $e$-$p$) momentum densities.

In this paper we demonstrate what one can get from momentum densities $\rho(\mathbf{p})$ derived from densities reconstructed from 2D ACAR and CPs (Chapter 4), showing results obtained for the FS studies (Chapter 5). List of abbreviations used in the paper is given in Chapter 4.

## 2. Positron annihilation and Compton scattering techniques.

There is a variety of experimental techniques measuring either directly FS or some quantities connected with FS. They can be divided into two groups: magnetic and non-magnetic methods. Magnetic methods (e.g. dHvA effect and resonance techniques) are connected with periodic oscillations of various physical properties (e.g. magnetic susceptibility) that depend on the electron energy. They allow to estimate only some quantities (e.g. area of extremal electron orbits) related only to FS (without visualization of investigated surface). Meanwhile, ACAR or Compton scattering spectra yield information on the shape of FS in an arbitrary point of the reciprocal space.

Positrons (with kinetic energy ~ 500 KeV) after implanting into the sample (mostly $^{22}$Na, $^{58}$Co and $^{64}$Cu are used) lose their kinetic energy and reach thermal equilibrium with the sample. During this process, if there is a low density region (as e.g. in defects where there is no positive atomic cores or molecular and ionic solids), positron can capture an electron forming like hydrogen a positronium atom. However, in the case of metallic samples free of defects one can assume that a positron annihilates from its ground Bloch state. Since the probability of emitting $n$ quanta γ is proportional to $(1/137)^n$, the most probable is the 2γ process (of course, in the case of antiparallel spins of annihilating $e$-$p$ pair) utilized in studying electronic structures of metals and their alloys (for more details see [3] and Chapter 11 in [2]).

Energy, momentum, mass and charge conservations cause that if the momentum of the $e$-$p$ pair is equal to zero ($|p| = 0$), 2γ rays are antiparallel ($\Theta = 0$), each one with momentum $mc$ ($m$ - electron mass, $c$ – light speed). When $|p| \neq 0$ we observe a distortion from the colinearity illustrated in Fig. 1.

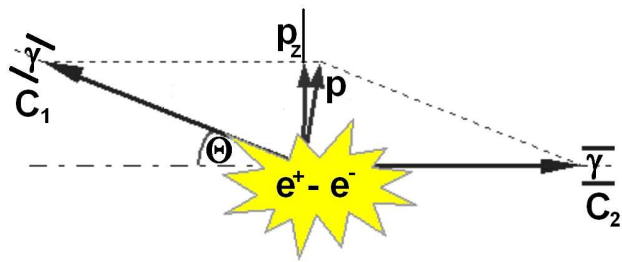

Fig.1. Geometry of 1D ACAR equipment. $C_1$ and $C_2$ - pair of counters being in coincidence; $p$ and $p_z$ momentum of the $e$-$p$ annihilating pair and its $z$ component in the laboratory frame[1].

---

[1] It is well known that directions [hkl] in the reciprocal space are defined by planes (hkl) in the real space being mutually perpendicular. Thus, putting an investigated monocrystalline sample to have its crystalographic plane {defined in the real (position) space by the Miller index (hkl)} parallel to counters $C_1$ and $C_2$ in their zero position ($\Theta = 0$), we define direction $z$ in the reciprocal (momentum) space by the same index [hkl].

Since $|p| \ll mc$, measured angles $\Theta$ are very small, changing between $0^0$ and $2^0$ where $1^0 \cong 17.5$ [mrad]. [mrad] denotes the momentum in the units $[10^{-3}mc=1]$, i.e. [mrad] = 0.137 [a.u.]$^{-1}$ (atomic units of momentum). So, e.g. electrons inside the central FS are observed for angles $\Theta < 0.268^0$ (such angle corresponds to $|p_F| = 0.75$ [a.u.]$^{-1}$). Since $p_z/mc = sin(\Theta)$, for such small angles $\Theta = p_z/mc$, i.e. the angular correlation of the 2γ rays reflects the momentum distribution of the annihilating *e-p* pair. In the case of measuring 1D correlations, one gets 1D ACAR spectrum $N(\Theta) = N(p_z) = \int_{-\infty}^{+\infty}\int_{-\infty}^{+\infty} \rho^{e-p}(\mathbf{p}) dp_x dp_y$, representing plane projections of the *e-p* density, $\rho^{e-p}(\mathbf{p})$. Since a positron is thermalized, $\rho^{e-p}(\mathbf{p})$ corresponds to the electron momentum density with its breaks at FS, "seen" by positrons.

Present experimental equipments contain two sets of counters (e.g. 70×70), which allows to measure $N(\Theta, \varphi) = N(p_z, p_y)$, i.e. 2D ACAR spectra representing line projections of $\rho^{e-p}(p)$.

In Fig. 2 we show 2D ACAR spectrum for the hexagonal alpha-quartz, attributable to the momentum distribution of a parapositronium. It illustrates that in ACAR experiment one obtains information function $\rho^{e-p}(p)$ in the extended momentum space where both central peak and Umklapp components around the reciprocal lattice vectors are clearly seen.

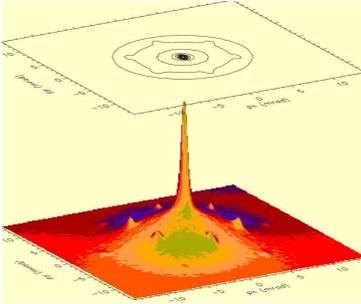

Fig. 2. 2D ACAR spectrum for the hexagonal alpha-quartz with the integration direction normal to the basal hexagonal plane [14].

During the Compton scattering the photon transfers a fraction of its energy to the electrons. The total kinetic energy of the system is unchanged, the number of interacting objects remains the same and there is no energy transfer to other forms (so some authors call it elastic scattering). However, because not only angle $\Theta \neq 0$ but also $\hbar\varpi \neq \hbar\varpi'$ (see Fig. 2), other authors use the term inelastic scattering.

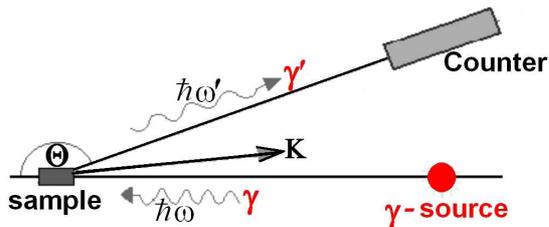

Fig. 3. Schematic diagram of measuring I($\omega'$) in the Compton scattering experiment with fixed angle $\Theta$ (usually about $165^0$).

I($\omega'$) reflects the momentum distribution of electrons having momenta $p_z$ (i.e. plane integral as in the case of 1D ACAR) where the direction $p_z$ is along a scattering vector $\mathbf{K}$ - for more details see Chapter 2 in [2].

In both measurements (ACAR and Compton scattering) one does not measure absolute values of densities $\rho(p)$. However, in the Compton scattering experiment the total integral of the electron momentum density should be equal to the number of electrons per unit cell - Compton scattering samples all electrons (valence and core) with the same probability. This is not the case for ACAR spectra where a positron (positive particle thus repelled from positive ions) favors regions outside the ionic cores, i.e. conduction electrons. Moreover, due to the *e-p* interaction, the electron density is enhanced by the positron, So, the total integral of the *e-p* momentum density over the whole *p* space is given by the number of electrons per unit cell "seen by a positron", i.e. spectra should be normalized to the inverse of the lifetime of a positron in the material.

Many-body effects in both experiments (influenced investigated function $\rho(p)$, though without changing FS) and ways of dealing with experimental data are described in Chapter 4.

### 3. Image reconstruction from projections.

Generally, functions $g$ and $\rho$ can be expanded into spherical harmonics $S_l$ defined on $R^N$:

$$g(r,\zeta) = \sum_l g_l(r) S_l(\zeta) \quad \text{and} \quad \rho(p,\omega) = \sum_l \rho_l(p) S_l(\omega). \quad (3)$$

Doing this, radial functions $g$ and $\rho$ are the Gegenbauer transform pair [8] where

$$\rho_l(p) \cong \frac{1}{p} \int_p^\infty g_l^{(2\mu+1)}(r) C_l^\mu(r/p) \left[(r/p)^2 - 1\right]^{\mu-1/2} dr. \quad (4)$$

$\mu = N/2 - 1$, $g^{(n)}$ denotes the *n*-th derivative of $g$ and $C_l^\mu$ are Gegenbauer polynomials. The singularity in Eq.(4) as well as an estimation of derivative of experimental quantities (burdened with statistical noise) makes its application in numerical calculations difficult. This can be circumvented if $g$ is expanded into such orthogonal polynomials that Eq.(4) is solved analytically giving $\rho_l(p)$ in terms of other polynomials. Moreover, such an expansion, having a similar effect as the mean-squares fitting procedure, essentially reduces the experimental noise when applied to real data. Here two sets of polynomials were proposed: the first found by Louis for *N*-dimensional space [15] in terms of Gegenbauer and Jacobi polynomials and the second one in terms of Hermitte and Laguerre polynomials, both of them derived earlier by Cormack [5] for *N*=2. Cormack's method (CM), adopted for symmetry systems [16], has been applied many times to reconstruct either *e-p* momentum densities from 2D ACAR data [17-32] or line dimensions of FS from CPs (conversion from 1D to 2D densities, i.e. from plane to line projections) [33-37]. The equivalent solution for *N* = 3 (plane integrals) in terms of Jacobi polynomials [38] up to now has been employed to reconstruct electron densities from CPs in Y [39], $Cu_{0.1}Al_{0.9}$ [40], and the shape-memory alloy $Ni_{0.62}Al_{0.38}$ [41].

Hermitte and Laguerre polynomials for $N=3$ were considered by Reiter and Silver [42] (see also Ref. [43]) and utilized to neutron scattering experiments [44, 45]. Such a solution was also found by Mijnarends [6] who (due to poor computer abilities in 1967) proposed another formula. Mijnarends' method, applied to both 1D ACAR data and Compton scattering profiles, has been used in years 1969 to 1975 – for more details see the overview paper [46].

In the Fourier transform (FT) techniques [8, 9, 47] one calculates the 1D FT of measured spectra $g$:
$$F(q,\zeta) = 2\int_0^\infty g(r,\zeta)\cos(2\pi rq)dr, \tag{5}$$

getting either 2D or 3D FT of $\rho(p)$, for $N=2$ and $N=3$ respectively. Next, the reconstructed density is evaluated from the inverse FT. For line projections it could be written in the form
$$\rho(p,\Theta) = \int_0^\pi W(p\cos(\Theta-\varphi),\varphi)d\varphi \quad \text{where} \quad W(s,\varphi) = 2\int_0^\infty F(q,\varphi)|q|\cos(2\pi sq)dq \tag{6}$$

Utilizing the convolution of filtering and sampling theorem, two integrals (Eq. (5) and this defined function $W$) can be reduced to the following summation [7]:
$$W(s_i,\varphi) = \frac{g(s_i,\varphi)}{4\Delta s} - \frac{1}{\pi^2\Delta s}\sum_{j=1}^n \frac{g(s_j,\varphi)}{(i-j)^2}, \tag{7}$$

over all $j$'s for which $i-j$ is odd, where $\Delta s$ denotes the distance between experimental points.

This method, named filtered back projection (FBP), was applied to first 2D ACAR measurements [48-54].

Contrary to medical investigations, for studying electronic densities it is sufficient to measure a few projections and introduce the angular interpolation (Eqs. (2-3)) either to experimental data or to the function $W(s,\varphi)$ [55]. Such a method, named modified FBP (MFBP), was applied (parallel with the CM) in the papers [21, 24-29].

There is also another possibility. The expansion of the FT, $F(q,\varphi)$, into the lattice harmonics eliminates the integration over $\varphi$ and expression for $\rho_n$ becomes (e.g. [47]):
$$\rho_n(p) = 2\pi(i)^n \int_0^1 F_n(q)|q|^{N-1} J_n(2\pi pq)dq, \tag{8}$$

where $J_n$ denotes the Bessel function of the first kind. Such a procedure, called Fourier-Bessel (F-B) method, was applied to reconstruct 2D densities from 2D ACAR data [56] and 3D densities from 1D Compton profiles [57-60]. However, because calculations of Bessel functions of a higher order make some difficulties, lately the direct FT (instead of F-B) algorithm [61, 62] has been used to both 2D ACAR [63-82] and 1D Compton profiles, reconstructing either fully 3D densities [83-89] or 2D ones [90-96]. Such techniques involving both fast FT algorithm and different ways of interpolations (instead of angular interpolation as used by us in MFBP) were elaborated by many authors, e.g. [97].

When measured 2D spectra are not collected in such a way that the reconstruction of a 3D density can be reduced to a set of reconstructions of 2D densities, one can use some technique for plane projections. Namely, for each 2*D* spectrum one estimates some number of 1*D* spectra $g(p_z)$ for different directions $p_z$. Next, densities are reconstructed from plane projections by applying either the F-B method (as proposed by Pecora [98] and applied in papers [99-104]) or another techniques as discussed in Ref. [105] where also various reconstruction algorithms for both line and plane projections are compared.

We have found about 100 papers where such techniques were applied to study fermiology via momentum densities reconstructed from 1D (or 2D) ACAR and 1D Compton scattering experimental spectra. All of them belong to the transform methods described above, except for the maximum entropy algorithm [106], applied to 1D CPs [107-109]. In the case of 1D spectra either fully 3D or 2D densities were reconstructed (this way of dealing with data is explained in the next Chapter). It is summarized in Table 1 where the following abbreviations are used:

CM – Cormack's method with Chebyshev polynomials; DFT – direct Fourier transform; FBP – filtered back projection [7] with using Eq. (7); F-B – Fourier Bessel; JP – Jacobi and HP- Hermite polynomials; ME – maximum entropy; MFBP – modified FBP; PM - Pecora method.

Tabele 1. Reconstruction techniques applied to 2D-ACAR and 1D-CP spectra

| Experiment ⇒ reconstruction | applied technique | References | When | how many |
|---|---|---|---|---|
| 2D ACAR ⇒ 3D densities | FBP | [48-54] | 1979/1989 | 7 |
| | PM | [99-104] | 1985/1993 | 6 |
| | CM | [17-32] | 1989/2007 | 16 |
| | F-B | [56] | 2006 | 1 |
| | DFT | [63-82] | 1989/2008 | 19 |
| | MFBP | [21, 24-29] | 2001/2007 | 7 |
| 1D CP ⇒ 3D densities | F-B | [57-60] | 1987/1999 | 4 |
| | JP | [39-41] | 2002/2006 | 3 |
| | DFT | [83-89] | 1993/2008 | 7 |
| | ME | [107-109] | 1995/2001 | 3 |
| 1D CP ⇒ 2D densities | CM | [33-37] | 1997/2007 | 5 |
| | HP | [43] | 1987 | 1 |
| | DFT | [90-96] | 2001/2006 | 7 |

In the paper the following abbreviations are also used: 2D (or 3D) – 2 (or 3) dimensional; ACAR - angular correlation of annihilation radiation; CP – Compton profile; dHvA - de Haas van Alphen; *e-e* – electron-electron; *e-p* – electron-positron; FS – Fermi surface; FT - Fourier transform; FP - full potential; LCW – (Lock–Crisp–West) authors paper [118]; LMTO-ASA – linear muffin tin orbital - atomic sphere approximation.

## 4. Analysis of experimental data.

Generally, there are two ways of dealing with ACAR and Compton scattering experimental data: $1^0$ - experimental profiles are compared with theoretical ones, calculated for some model $\rho(\boldsymbol{p})$, based on band structure calculations with including (or not in the case of using independent particle model, IPM) many-body correlations;  $2^0$ - 3D densities $\rho(\boldsymbol{p})$ are reconstructed directly from experimental profiles.  Of course, the best solution is when these two ways are applied simultaneously, i.e. 3D densities $\rho(\boldsymbol{p})$ are reconstructed also from a set of theoretical spectra (the same as experimental ones). Such a procedure is usually used in an analysis of 1D-CPs, where (comparing to 2D-ACAR spectra) it is much easier to calculate theoretical profiles though it is much more difficult to reconstruct properly 3D $\rho(\boldsymbol{p})$.

**4.1. many-body effects**

Because in both experiments there are dynamic processes, one should include into the theory, beyond band structure calculations, many-body effects.  In the Compton scattering there are *e-e* correlations which (in the simplest approximation) are described by the isotropic Lam-Platzman correction [110]: diminishing of densities in the low-momentum region, smearing around FS, and so-called many-body tail above the Fermi momentum $p_F$. In the case of positron annihilation one has to deal with a system consisting of many electrons and one positron moving in a crystal lattice and interacting with one another. So, an ideal theoretical description of the *e-p* annihilation in metallic materials should include (beyond the IPM where electron and positron wavefunctions are based on the band structure calculations) both *e-p* and *e-e* correlations inside the periodic lattice potential. It is evident that such a problem can be solved only with rude approximations. All theories (except for Arponen and Pajanne's theory [111] where a positron in an interacting electron gas is considered)  are based on the result of Carbotte and Kahana [112] where *e-p* pair, seen from outside, is a neutral quantity with a strongly reduced coupling to its environment. The remaining influence of many body effects comes only from the static part of the *e-p* interaction (dynamic parts of the *e-p* and *e-e* correlations cancel themselves) – for review see e.g. [113, 114]. However, as shown lately on the example of Y (by simultaneous analysis of high-resolution CPs and 2D ACAR spectra [39]) and for Mg (see Fig. 4) this is not true, i.e. there are the same dynamic *e-e* correlations in both experiments.

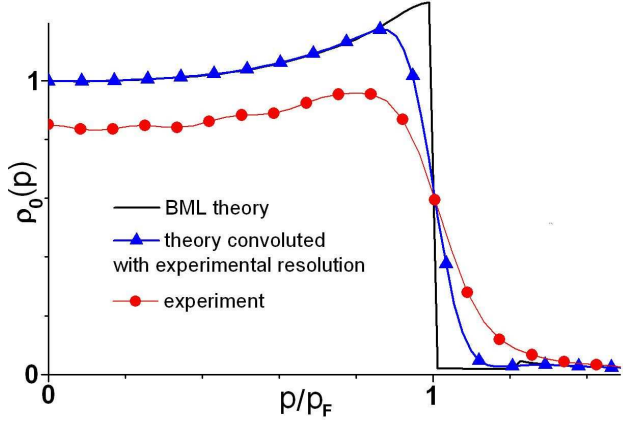

Fig. 4. The isotropic average of the momentum density in Mg as a function of $p$ in units of the Fermi momentum $p_F$. "Experimental" and theoretical densities are normalized to the experimental total annihilation rate – more details in [113].

Theoretical results were based on electron and positron wave functions obtained by the full potential linearized augmented plane wave method within the local density approximation and including scalar-relativistic effects. In order to describe the *e-p* interaction the Bloch modified ladder approach [115] was applied where, contrary to other theories, the *e-p* correlations are introduced via the periodic lattice potential. After normalizing densities to the experimental total annihilation rate (the inverse of the experimental life-time) we were able to observe an effect typical of dynamic *e-e* correlations [110]. In Mg we observe also the Kahana-like enhancement which monotonously increases densities with increasing a momentum. Due to the Pauli principle, in the case of an electron gas where all states inside FS are fully occupied, perturbated states can appear only for ***p*** $> p_F$. Consequently, since scattering is the most probable for electron states close to FS, the enhancement has maximal value at FS. However, the higher lattice effects are, the weaker is the Kahana-like momentum-dependence of the enhancement (got within the Bloch modified ladder approach for Mg but not for more complicated materials as e.g. Y [113]).

### 4.2. From 1D-Compton profiles to 2D densities

In principle, reconstruction of 3D densities from plane projections demands a large number of profiles. Thus, for 1D data, it is reasonable to reconstruct 2D density [33], defined as $\rho^L(p_z, p_y) = \int_{-\infty}^{\infty} \rho(\mathbf{p}) dp_x$, where 1D profile, being a plane integral of 3D density $\rho(p)$, is treated as a line integral of $\rho^J$. We demonstrate this on the example of $\rho^J$ in Be, reconstructed via the CM from both experimental and theoretical CPs [35]. Due to the hexagonal symmetry, in this case despite the momentum density is highly anisotropic, merely two Compton profiles (measured [116] with $p_z$ along ΓM and ΓK) were sufficient to reproduce the main features of $\rho^J(p_z,p_y)$ (with $p_x$ along the hexagonal $c$-axis) displayed in Fig. 5.

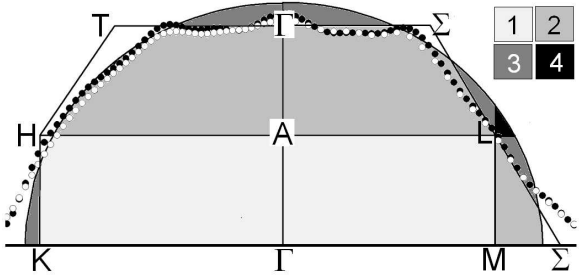

Fig. 5. $\rho^J$ in Be for momenta along ΓK and ΓM, reconstructed from two CPs. Theory and experiment are marked by solid and open symbols, respectively. Full line draws free electron FS in the four bands in the extended zone.

Theoretical profiles were calculated within the self-consistent band structure theory [117], including the Lam-Platzman correction [110] and the experimental resolution. Compared to the free-electron model (marked by full line in Fig. 5), they correspond to the following feature of FS: a) no holes around the $H$ point either in the 1st or 2nd band (fully occupied 1st zone and no holes in $2^{nd}$ zone on the plane $AHL$; b) very small holes around $\Sigma$ and reduced holes around $T$ in the $2^{nd}$ zone, compared to the free-electron model; c) no electrons around $\Gamma$ in the $3^{rd}$ band; d) no electrons around $L$ either in the $3^{rd}$ or $4^{th}$ bands; and e) cigars in the $3^{rd}$ zone around $K$ are larger than for the free-electron model with their height close to $|KH|$.

It is seen that absolute densities are not reproduced exactly, e.g. a small electron-like lens is observed at $\boldsymbol{p} = 0$ – the most probably an artifact originating from the fact that a density is reconstructed from only two projections and there are always high reconstruction errors around $\boldsymbol{p} = 0$ (the more so as reconstructed theoretical $\rho^J$ show similar effect). Moreover, in the analysis of densities in the $\boldsymbol{p}$ space one should bear in mind that $\rho^J$ cannot be directly identified with line dimensions of FS because in real metals $\rho(\boldsymbol{p}) < 1$ also in the central FS (in the $\boldsymbol{p}$-space there are both central and Umklapp surfaces). However, reconstructed densities clearly show the lack of electrons around the $L$ point in the 3rd and 4th bands and the shape of FS on the $\Gamma MA$ plane close to the double Brillouin zone boundaries.

**4.3. Conversion from extended $p$ to reduced $k$ momentum space.**

Taking into account the complexity of a many-body problem of electrons and a positron moving within a lattice potential, almost all interpretations of experimental ACAR data are performed only from the point of view of the FS studies. In order to obtain the contour of FS, the best way is to fold $\rho(p)$ from the extended zone $p$ into the reduced momentum space $k$ via the LCW-folding [118] to obtain $\rho(\mathbf{k}) = \sum_{\mathbf{G}} \rho(\mathbf{p}=\mathbf{k}+\mathbf{G}) = \sum_{l} n_l(k)$, where $k$ denotes vectors in the first Brillouin zone and the summation is performed over the reciprocal lattice vectors $G$. Due to such a procedure lattice effects are re-emphasised and $\rho(k)$ shows the sum, over the bands $l$, of the occupation numbers $n_l = \{1,0\}$ for filled and empty states, respectively. Of course, a positron wavefunction and many-body correlations somehow affect the determination of FS (changing densities but not $k_F$) and $n_l$ must be modified by some function $f_j(k)$ which contains both many-body and positron wave function effects. In most cases the FS breaks are sufficiently intense to reveal the FS topology [119, 120], although, as shown lately for $UGa_3$ [121], sometimes a presence of the positron does not allow for a precise analysis of experimental $\rho(k)$ without corresponding theoretical calculations.

It is not a case for electronic densities studied by the Compton scattering (where $f_j(k)=1$), demonstrated in Fig. 6 on the example of densities reconstructed from CPs in $Cu_{0.9}Al_{0.1}$ [40].

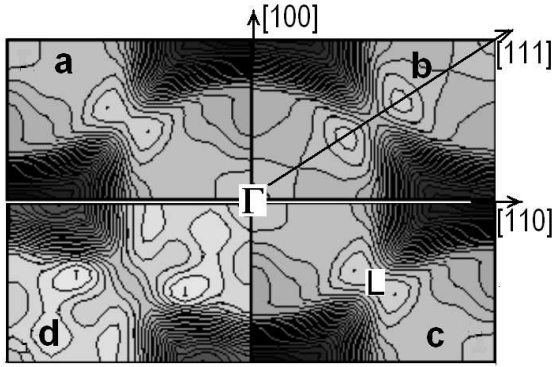

Fig. 6. Folded momentum density $\rho(k)$ in $Cu_{0.9}Al_{0.1}$ in the (110) plane (in the repeated zone scheme) obtained from 9 theoretical (part a and c) and experimental (parts d and b) CPs. Part (b) show experimental densities after subtracting $e$-$e$ correlation effects (experimental $\rho_0$ is replaced by theoretical one).

The folded $\rho(k)$ derived from the theoretical CPs (computed within the fully self-consistent band structure caculations [122]) and from experimental CPs reasonably display signatures of the well-known FS of Cu. The main discrepancy between theoretical and experimental $\rho(k)$ is observed along the [111] direction around the neck. This discrepancy is much higher than the corresponding error connected with the experimental noise. The most probably it is connected with the e-e correlation effect which is demonstrated in the part (b) of Fig. 6 where experimental $\rho_0(p)$ displayed in the part (d) (it contains $e$-$e$ correlations) was replaced by theoretical $\rho_0(p)$ (without correlations).

In Fig. 7 we show folded $e$-$p$ densities, $\rho^{e-p}(k)$, for $ErGa_3$ (in paramagnetic phase) [25]. In this case $f_j(k) \neq 1$, i.e. different elements of FS are probed by a positron with different probabilities which is more clearly illustrated in Fig. 8

Recently, FS of paramagnetic ErGa$_3$ has also been probed by three new codes with the full potential (FP) instead of atomic sphere approximation (ASA): FP linear muffin tin orbital (FP-LMTO), FP linear augmented plane wave and FP local orbitals methods [32]. Surprisingly, none of these codes is able to reproduce the experimental results which agree very well (as shown in the Fig. 7) with the former LMTO-ASA band structure results [123]. The conclusion drawn in the paper [32] is the following "it can be an evidence of some failure in construction of an atomic potential or an inefficient choice of internal parameters, presumably the linearization energy $E_{vl}$". Standard FP codes (applied also in [32]) use $E_{vl}$ as the center of gravity of an occupied $l$ band. However, as it was shown by Skriver [124] and applied to previous LMTO-ASA calculations for ErGa$_3$ [121], another possible choice, $E_{vl} = E_F$, gives more accurate FS.

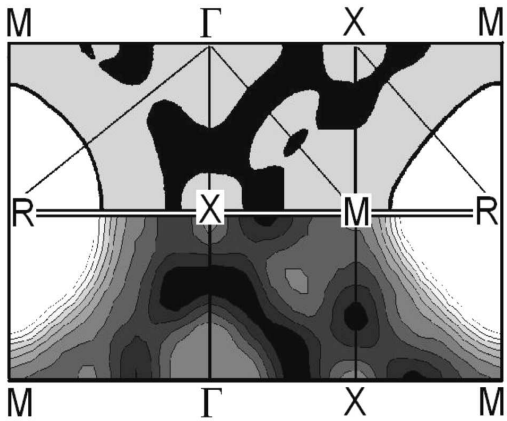

Fig. 7. Densities $\rho^{e-p}(\mathbf{k})$ in ErGa$_3$ on three high symmetry planes, reconstructed by CM, compared to the FS sections calculated via LMTO-ASA [123]. The white region centered at the R point contains the occupied states from the 7$^{th}$ valence band and the black area denotes unoccupied states from the 6$^{th}$ band.

Fig. 8. $\rho^{e-p}(\mathbf{k})$ in ErGa$_3$ (reconstructed from 2D ACAR spectra and shown in Fig. 7) along some high symmetry lines, in the units when $\rho(\mathbf{p})$ for $\mathbf{p} = 0$ is normalized to the unity.

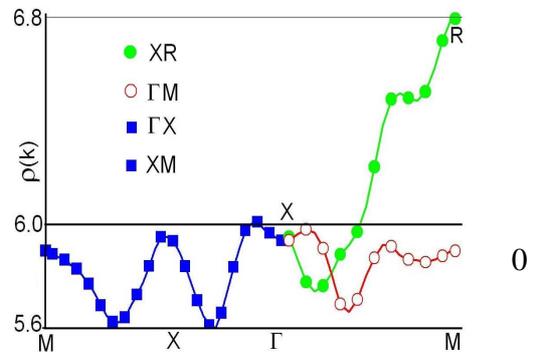

From Fig. 8 it is clear that the jump of densities between electrons in the 5$^{th}$ and 6$^{th}$ band (about 0.4) is two times lower than that between 6$^{th}$ and 7$^{th}$ bands. Thus, one can conclude that electrons in the 7$^{th}$ band around the R point must be "more free" (positron sees them with higher probability), i.e. they are mostly $s$-like while these in the lower bands mostly $d$-like. So, it shows that which in principle could be a disadvantage of the ACAR experiments in comparison with the Compton scattering technique (probing directly the electron densities) turns out to be advantageous. Namely, since the positrons are repelled from positive ions, it is possible to infer from ACAR data some information on the degree of the electron localization, as e.g. we showed here for ErGa$_3$.

Reconstructed densities, shown in figures 7 and 8, were "filtered" by imposing on the densities, reconstructed independently on the parallel planes (001), the symmetry requirement $\rho(p_x, p_y, p_z) = \rho(p_y, p_z, p_x) = \rho(p_x, p_z, p_y)$. Such a requirement follows from the fact that for the cubic structures non-equivalent fraction of the Brillouin zone is equal to 1/48, instead of 1/16 as for other structures with one *4*-fold main rotation axis. Such a procedure (imposed on densities either in *p* or *k* space) not only reduces an influence of the experimental noise but also allows to reconstruct densities from a smaller number of projections [125].

The equivalent procedure to the LCW folding is Schülke's method [126] (as shown explicity by Shiotani [127] ) where $\rho(k)$ is calculated from FT of CPs, so-called *B*(*r*) -function.

### 4.4. Analysis in extended *p* space.

Lately, we have proposed another filtering algorithm [29], based on a description of densities in the 3D space by the lattice harmonics, applying it to 2D ACAR data in $LaB_6$ [29]. Thanks to this method we reproduced a small element of FS (electron pocket in the 15th band along ΓM line – see Fig. 9) observed also in the dHvA experiments [128-130]. It had not been reproduced before by other reconstruction techniques applied to the same experimental data [24] as well as by the analysis of 2D ACAR data in the *k* space [131]. Here we would like to point out that results of the latest band structure calculations in $LaB_6$ [132] (contrary to the previous ones [133-135]) also show this element.

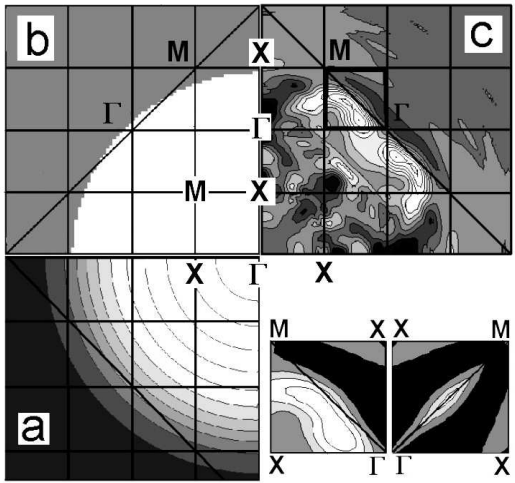

Fig. 9. Densities on the plane (001) in $LaB_6$ reconstructed from three deconvoluted 2D ACAR spectra: the isotropic average of densities, $\rho_0(p)$ - part (a); free-electron sphere containing 27 electrons - part (b); $\rho(p)$-$\rho_0(p)$ - part (c). All parts are drawn with some Brillouin zone boundaries.

Sometimes, the knowledge of $\rho(p)$ in the whole *p* space allows to extract dimensions of FS's in different bands via so-called symmetry selection rules [136]. A *k*-space analysis of 2D ACAR data in yttrium [19] exploited the near coincidence of the 3$^{rd}$ and 4$^{th}$ band surfaces on the *KMLH* Brillouin zone face. Authors obtained the shape and size of the so-called *webbing* feature at that face, but nevertheless, by analysis in the *k* space it was not possible to get information on the individual surfaces. Our further interpretation of the same 2D ACAR data but in terms of $\rho^{e-p}(\boldsymbol{p})$ with the knowledge of the theoretical densities and the effects of symmetry selection rules, has allowed us to separate two hole FS in the 3$^{rd}$ and 4$^{th}$ bands and to establish some Fermi momenta for each of them [137, 39].

## 5. Summary.

Results of studying electronic structure of metallic materials via momentum densities reconstructed from positron annihilation and Compton scattering techniques are summarized in Tables 2 - 4 with including investigations of spin densities by measuring magnetic profiles. There were also studies (not presented here but shown e.g. in Refs. [2 - 4]) where information on FS was derived directly from experimental profiles.

Tabele 2. FS obtained from 3D *e-p* momentum densities reconstructed from 2D ACAR data.

| material | results concerning FS | Ref. |
|---|---|---|
| Al | FS on (100) and (110) planes | [68] |
| Al | hole surface around Γ, electron surfaces along lines WUW and WKW. No gap between electron surfaces at W interpreted as effect of the experimental resolution | [80] |
| Mg | quantitative information on distortion of FS from sphericity | [23] |
| Cd | lack of the 3$^{rd}$ & 4$^{th}$ zone electrons around L; reduction of hole monster to 6 separate hole surfaces nearby K | [23] |
| Cu | FS on (100) and (110) planes | [49, 71, 82] |
| Cr in 323 K & 30 K | small differences at R & Γ points and along Λ & Σ lines in paramagnetic & antiferromagnetic states | [75] |
| Cr (323 K), Mo & W (30 K) | Γ-centered electron surface and hole surfaces at H and N – the later only in Mo and Cr | [67, 77] |
| V, Nb & Ta | Γ-centered hole octahedron, multiplay connected jungle-gym arms and N-centered hole ellipsoids | [76] |
| V | ratio of NP/NH of semiaxes of N-centered ellipsoidal hole FS equals 1.36 | [99, 100] |
| Co | polarized 3D densities (and occupation numbers) in agreement with theory except for those around the K point. | [72] |
| Ti & Zr | electron surface at H and hole surface along ΓA. | [64, 66] |
| Y | size & shape of FS in the vicinity of the A-L-H plane known as the webbing - first experimental direct observation of such phenomena | [19] |
| Gd | FS (for spin polarized spectra) in agreement with dHvA results [138] | [50] |
| Gd | compared to [138]: similarity of FS on the ΓALM plane but not on ΓAHK (around K). | [51, 54] |
| $Gd_{0.62}Y_{0.38}$ | reconstructed FS and corresponding nesting vector, in agreement with the period of helicity | [20] |
| $TmGa_3$ | nesting of FS along the [110] direction in the paramagnetic phase consistent with the antiferromagnetic structure | [21] |
| $ErGa_3$ | nesting of FS (in paramagnetic phase) consistent with the modulated antiferromagnetic structure | [25, 32] |
| $CeIn_3$ | good agreement with theory when 4f electrons are fully localized | [27] |
| $UGa_3$ | good agreement with theory when 5f electrons are itinerant | [28] |
| $LaB_6$ | all elements of FS, together with small electron pocket in 15$^{th}$ band | [29] |
| $CeB_6$, $LaB_6$, $PrB_6$, $NiB_6$ | the main structure of $\rho(k)$ in agreement with the FLAPW band structure | [63] |
| $TiBe_2$ | good agreement with LMTO results | [101] |
| $LaRu_2Si_2$ & $CeRu_2Si_2$ | similar results for heavy-fermion $CeRu_2Si_2$ (above $T_K$) and non-f-electron $LaRu_2Si_2$. Better agreement with theory for $CeRu_2Si_2$ | [26] |
| $ZrZn_2$ | in the paramagnetic state, flat FS in 27 band (pillows), excellent candidate for FFLO superconducting state | [30] |
| Cr, $Cr_{0.7}V_{0.3}$, $Cr_{0.85}Mo_{0.15}$ | evolution of period of the oscillatory exchange coupling directly connected with the evolution of N-hole ellipsoid FS | [31] |
| $V_3Si$ | two nested cylindrical surfaces along the zone edges forming hole surface around R; electrons at the X point | [48, 53] |
| $Li_{1-x}Mg_x$ ; x=0, 0.28, 0.4, 0.6 | detailed studies of critical concentration at which FS sticks the Brillouin zone along [110]. N-hole pockets and hole octahedron at H are observed | [102, 103] |
| $Cu_{1-x}Pd_x$ | evolution of FS with x; the strongest nesting for x=0.28 | [22] |
| $La_{0.91}Sr_{0.09}CuO_4$ & $La_2CuO_4$ | FS is 2D and consists of electron pillar along ΓZ and two kinds of hole pillars at X and N. | [65] |
| GaSb & InP | good agreement with the Jones zone scheme; some distortions interpreted as interference effect of wavefunction | [69] & [70] |
| $Nd_{2-x}Ce_xCuO_4$ $Pr_{2x}Ce_xCuO_4$ | flat occupation numbers as in semiconductors/or insulators for x=0 & x=0.04 (without & with carrier doping) | [73] |
| $Bi_2Sr_2CaCu_2O_8$ | flat occupation numbers as in semiconductors/or insulators in the superconducting and normal state | [74] |
| 2H-$NbSe_2$ | open cylindrical hole surface along ΓA; second hole surface along KH not found. | [79] |
| β'-AgZn | FS similar as in β'-CuZn - 1$^{st}$ zone hole octahedron at R & 2$^{nd}$ zone electron surface | [81] |

Tabele 3. Information on FS via 3D electron momentum densities reconstructed from 1D CPs

| | | |
|---|---|---|
| Li | anisotropy of FS $(k_{[110]} - k_{[100]})/k_F^{free} \cong 3,6\%$ | [59, 84, 107] |
| Ni | magnetic profiles in ferromagnetic phase, good agreement | [89] |
|   | with band structure results |   |
| Fe | shown that FLAPW theory slightly underestimates negative spin polarization of $s$, $p$-like electrons in the 1$^{st}$ Brillouin zone | [83] |
| Fe, $Fe_3Si$ & Heusler alloy $Cu_2MnAl$ | magnetic CPs: negative polarization of conducting electrons for $Fe_3Si$ and Fe are similar to each other while for $Cu_2MnAl$ it is much smaller, if exists at all | [108] |
| $CoSi_2$ | FS is not drawn | [109] |
| Y | good agreement with band structure and 2D ACAR data [19]; line dimensions of FS | [39] |
| $Cu_{0.9}Al_{0.1}$ | FS similar to that of Cu | [40] |
| $Ni_{0.62}Al_{0.38}$ | nesting vector = 0.18* [1,1,0]($2\pi/a$) | [41] |
| $Li_{1-x}Mg_x$ | for x > 0.13 neck along [110] | [60] |
| GaAs | good agreement with theory | [58] |
| TiNi | nesting of FS | [86] |
| β-$PdH_{0.84}$ | FS almost spherical with the neck in [111] – like in Cu | [87] |
| Cu-27.5 at.% Pd disordered alloy | Fermi momenta on two high symmetry planes | [85] |
| $Ba_{1-x}K_xBiO_3$ | For insulating phase (x=0.1) experimental results very similar to the theory for x=0. For metallic sample (x=0.37) results similar to the theory but still show unusual feature due to the FS nesting | [88] |

Tabele 4. FS obtained from 2D electron momentum densities reconstructed from 1D CPs

|   |   |   |
|---|---|---|
| Cr | electron jack at Γ, hole octahedral at H & hole ellipsoids at N | [34, 90] |
| Be | absence of FS around Γ and L (3$^{rd}$ & 4$^{th}$ bands) | [35] |
| $Na_xCoC_2$; x=0.74; 0.51 & 0.38 | small elliptical hole pocket for low concentration Na and in hydrated phase (more details below) | [36] |
| Al-3 at.% Li disorder alloy | good agreement with the theory | [91] |
| $Ba_{1-x}K_xBiO_3$; x=0.13 & 0.39 | metallic phase: discrepancies with theory around L interpreted as gap opening; insulating phase – filled polyhedral Brillouin zone and perfect nesting of FS | [94] |
| $Sr_2RuO_4$ at 20 K & RT | results for 20 K in agreement with FLAPW, thermal behaviour not understood | [95] |
| bilayer manganite | the coexistence of polaronic and band states in the FM phase | [96] |
| η-$Mo_4O_{11}$ | good agreement with theory (tight-binding method) except for too much smeared Y-Z hole channel interpreted as strong e-e correlation effect | [93] |

There are no doubts that the knowledge of the FS topology is a crucial point in understanding electronic properties of quantum systems. The principal reason for the FS importance lies in the fact that, due to the Pauli principle, only electrons at FS can respond to external fields.

So, there is a strong connection between the FS topology and various exotic phenomena, as e.g. magnetism in the heavy rare earths [139], spin density waves or other phenomena that accompany quantum criticality [140] and unconventional superconductivity [141].

Comparing to magnetic methods (like dHvA) which measure only some parameters of extremal electronic orbits without any visualization of FS, 3D momentum densities reconstructed from either ACAR or Compton scattering spectra yield information on the shape of FS in the arbitrary point of the reciprocal *k*-space. Moreover, they contain also information on the Umklapp components of the electron wavefunctions. Additionally, by measuring magnetic profiles, one can get information on spin densities.

Magnetic techniques of FS studies require both low temperatures and very pure samples. There is no temperature restrictions in the case of positron annihilation or Compton scattering measurements which allow to study materials in various temperatures, thus also in various physical phases. As concerns studying an electronic structure of alloys and metallic compounds with structural disorder the best is the Compton scattering techniques because positrons are trapped by defects. Of course, this technique has also some limitations – first of all a proper reconstruction of 3D densities from plane projections is difficult and requires many projections measured for very particular scattering directions, determined by special directions in the Brillouin zone [142]. However, here one restricts oneself to reconstruction of 2D densities where only a few projections are needed, as e.g. was done for Be (FS was derived from only 2 projections - see Fig. 5) or lately for hydrated sodium cobalt oxides [36]. From five 1D CP in $Na_xCoC_2$ and $Na_xCoC_2 \bullet 1.3H_2O$, authors showed that there are small elliptical hole pocket in FS for low concentration Na and in the hydrated phase. These pockets (crucial in model of spin-fluctuation-mediated superconductivity observed at 5 K in hydrated sodium cobalt oxides) were seen in Shubnikov-de Haas oscillations [143], phonon softening [144] but not in any of ARPES measurements [145-148]. There could be various reasons why these pockets are not observed in the ARPES experiments: surface sensitivity, including possible surface relaxations (as studied in $Na_xCoO_2 \cdot yH_2O$ [149]), matrix elements or a destruction of small FS elements by Na disorder [150].

Compton scattering and the allied techniques such as positron annihilation and ARPES (with much better resolution than typical ACAR machine and high-resolution CPs but having also many restrictions) are well described in Chapter 11 in [2]. Some questions connected with densities reconstructed from projections are discussed in the following papers: $1^0$ - an influence of the statistical noise on reconstructed densities [151 - 153]; $2^0$ – consistency conditions (projections of the same density must be interdependent), automatically imposed on experimental data via the reconstruction procedure, reduces some part of experimental noise [154]; $3^0$ - projections which should be measured in order to reconstruct densities properly [6, 142, 152, 153]; $4^0$ - efficiency of reconstruction techniques [105].

**Acknowledgements.** We are very grateful to Dr M. Biasini for making available his experimental 2D ACAR spectra for the hexagonal α-quartz, shown in Fig. 2, and to Prof. M.J. Cooper, Dr M. Samsel-Czekała, Prof. N. Shiotani, Dr Vyacheslav Fil and Dr E. Zukowski for helpful remarks.